\documentclass{article}

\usepackage{arxiv}
\usepackage[english]{babel}

\usepackage[utf8]{inputenc} 
\usepackage[T1]{fontenc}    
\usepackage{hyperref}       
\usepackage{url}            
\usepackage{booktabs}       
\usepackage{amsfonts}       
\usepackage{nicefrac}       
\usepackage{microtype}      
\usepackage{lipsum}
\usepackage{graphicx}
\usepackage{float}

\title{Automatic Detection of COVID-19 Cases on X-ray images Using Convolutional Neural Networks}

\author{
   Lucas P. Soares \thanks{\url{https://lpsmlgeo.github.io/}} \\
  MSc. student in Mineral Resources and Environment \\
  Institute of Geosciences  \\
  University of São Paulo (USP)\\
  São Paulo, Brazil \\
  \texttt{lpsoares@usp.br} \\
   \And
 César P. Soares \thanks{\url{https://cpscesar-en.github.io/}} \\
  Ph.D. candidate in Global Health and Sustainability \\
  Faculty of Public Health\\
 University of São Paulo (USP)\\
  São Paulo, Brazil \\
  \texttt{cpscesar@usp.br } \\
}

\begin{document}
\maketitle

\begin{abstract}
In recent months the world has been surprised by the rapid advance of COVID-19. In order to face this disease and minimize its socio-economic impacts, in addition to surveillance and treatment, diagnosis is a crucial procedure. However, the realization of this is hampered by the delay and the limited access to laboratory tests, demanding new strategies to carry out case triage. In this scenario, deep learning models are being proposed as a possible option to assist the diagnostic process based on chest X-ray and computed tomography images. Therefore, this research aims to automate the process of detecting COVID-19 cases from chest images, using convolutional neural networks (CNN) through deep learning techniques. The results can contribute to expand access to other forms of detection of COVID-19 and to speed up the process of identifying this disease. All databases used, the codes built, and the results obtained from the models' training are available for open access. This action facilitates the involvement of other researchers in enhancing these models since this can contribute to the improvement of results and, consequently, the progress in confronting COVID-19.

\end{abstract}


\section{Introduction}
In recent months the world has been surprised by the rapid advance of SARS-CoV-2 (COVID-19). Given the widespread of this virus on all continents, the World Health Organization (WHO) has declared that we are experiencing a pandemic of this disease.

To face this disease and minimize its socio-economic impacts, not only the surveillance and treatment are essential, but the diagnosis is also presented as a crucial procedure \cite{binnicker2020emergence}. However, the realization of this is hindered by the delay and limited access to laboratory tests available to detect COVID-19\footnote{ The Reverse-Transcriptase Polymerase Chain Reaction (RT-PCR) is considered the standard reference method for making the diagnosis of COVID-19 infection \cite{columbus20202019}.} \cite{world2020laboratory}, thus requiring new strategies to carry out case screening.

Studies in the area have pointed to the existence of specific indicators on the chest radiography of individuals infected with the SARS-CoV-2 virus \cite{ng2020imaging}. This fact would allow the use of these images in the diagnostic process of COVID-19 \cite{ai2020correlation}, expanding access to other forms of detecting the disease and accelerating its identification process.

In this scenario, deep learning models are proposed as a possible option to assist the diagnostic process. In particular, this technique arises intending to make the detection of cases of COVID-19 automatic from chest images - X-ray and computed tomography \cite{gozes2020rapid, xu2002deep, wang2020covid}.

This research proposes the training of models, using machine learning, to accurately detect the presence of COVID-19 from chest radiographs. Since the dataset used in this research has only 175 images of the positive class (COVID-19 chest X-rays), the authors trained the models using the transfer learning technique.

Complementarily, but not least, all the codes used to train the models and the data are available on GitHub\footnote{\url{https://deepdados-en.github.io/}}. This action aims to facilitate the involvement of other researchers in the process of enhancing these models, as this can contribute to the improvement of results and, consequently, the progress in facing COVID-19\footnote{Information on the databases used in the section referring to the methodology.}.

\section{Objectives}

Automate the process of detecting COVID-19 cases from chest radiograph images, using convolutional neural networks (CNN) through deep learning techniques.

\section{Methodology}

Deep learning is a machine learning technique that, through deep neural networks, seeks to discover autonomously - that is, without explicit programming - rules and parameters of a data set, in order to provide adequate representation for a particular problem. The term "deep learning" derives from a large number of hidden layers between the input and output layers of the neural network.

In this work, the convolution neural networks were trained using the supervised learning process \cite{lecun1995convolutional}. Therefore, the labels and the images served as input to the network, intending to minimize the loss function. This last one measures how far the prediction is from the expected output.

The models were trained with three different architectures: Xception \cite{chollet2017xception}, Residual Networks (ResNet) \cite{he2016deep}, and VGG-16 \cite{simonyan2014very}. Only the fully connected layers were optimized during the training process. The convolutional layers' weights were loaded from the ImageNet\cite{deng2009imagenet} dataset thought the transfer learning technique. The models were trained for 80 epochs with a learning rate of 0.001 and a batch size of 15. Categorical cross-entropy and Adam was used as loss and optimization function, respectively.

The images used to train the models had a size of 237 x 237 pixels and three colors channel (RGB). The training used 10\% of the images for validation and 10\% to test. Models were evaluated based on the accuracy metric. The results were plotted on a confusion matrix and analyzed using a class activation map (Grad-Cam) and by visual inspection. The models were trained on Google Colaboratory Virtual Environment using TensorFlow \cite{tensorflow2015-whitepaper} and Keras\cite{chollet2015keras} Python libraries.

The construction of the database aimed to maximize the number of training images. For this, two different sources were used: one containing X-ray images of the chest of individuals infected with COVID-19 (n = 175) \cite{cohen2020covid} \footnote{\url{https://github.com/ieee8023/covid-chestxray-dataset }} and the other, with the same type of image, but of lungs of individuals without any infection (n = 100) and with infections related to other virus and bacteria (n = 100) \cite{kermany2018labeled}\footnote{https://data.mendeley.com/datasets/rscbjbr9sj/2}. These datasets were used in this research because of the possibility of free access to the image banks.
\section{Results}

Each trained model's results were evaluated through the plots of the accuracy and loss function history of the training and validation data (figure \ref{fig:loss}).
{\setlength\intextsep{0pt}
\begin{figure} [H]
  \centering
  \includegraphics[width=450pt, keepaspectratio]{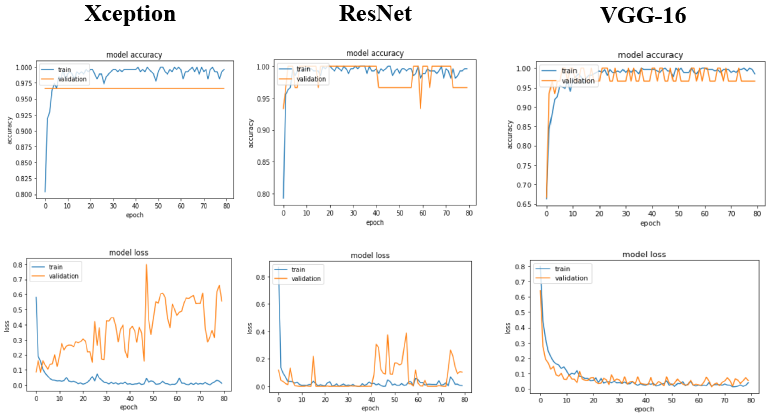}
  \caption{Accuracy and Loss plots from the training process.}
  \label{fig:loss}
\end{figure}
}

The Xception and ResNet models' graphs showed that although the accuracy was 95,9\%, 94,6\%, respectively, the models overfitted. This result can be inferred because the validation loss does not decrease according to the training loss. Nevertheless, the VGG-16 model graphics shows that there was less overfitting since the training and validation lines approached, and the model had the highest accuracy: 97,3\%, meaning that it classified correctly 97,3\% of the images used in the validation. Therefore, from the results, it is possible to notice that the VGG-16 model presented greater consistency and accuracy to classify the lungs' images.

The generalization capacity of the VGG-16 model was evaluated on a test set with 75 images, that were not used during the training process and randomly sampled from the dataset. The results were plotted in a confusion matrix.  
\vfill
{\setlength\intextsep{0pt}
\begin{figure}[H]
  \centering
  \includegraphics[width=350pt, keepaspectratio]{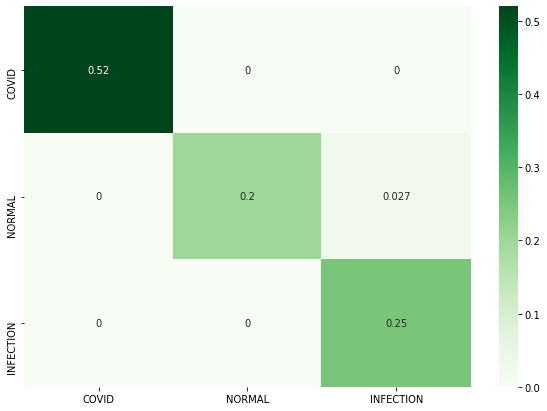}
  \caption{Confusion matrix - Network VGG-16.}
  \label{fig:matrix}
\end{figure}
}
\vfill
The matrix (figure \ref{fig:matrix}) pointed out that the accuracy value was 97,3\%. That is, among the 75 images used for the test, the model correctly classified 73. Accurately, the model correctly classified 100\% (n = 39) of the images related to COVID-19, 88,2\% (n = 17) of the images of normal lungs, and 100\% (n = 19) of the images of lungs with other infections. Regarding the errors, the model classified 2,7\% (n = 2) of the images referring to normal lungs as other infections.
{\setlength\intextsep{0pt}
\begin{figure} [H]
  \centering
  \includegraphics[width=\textwidth, keepaspectratio]{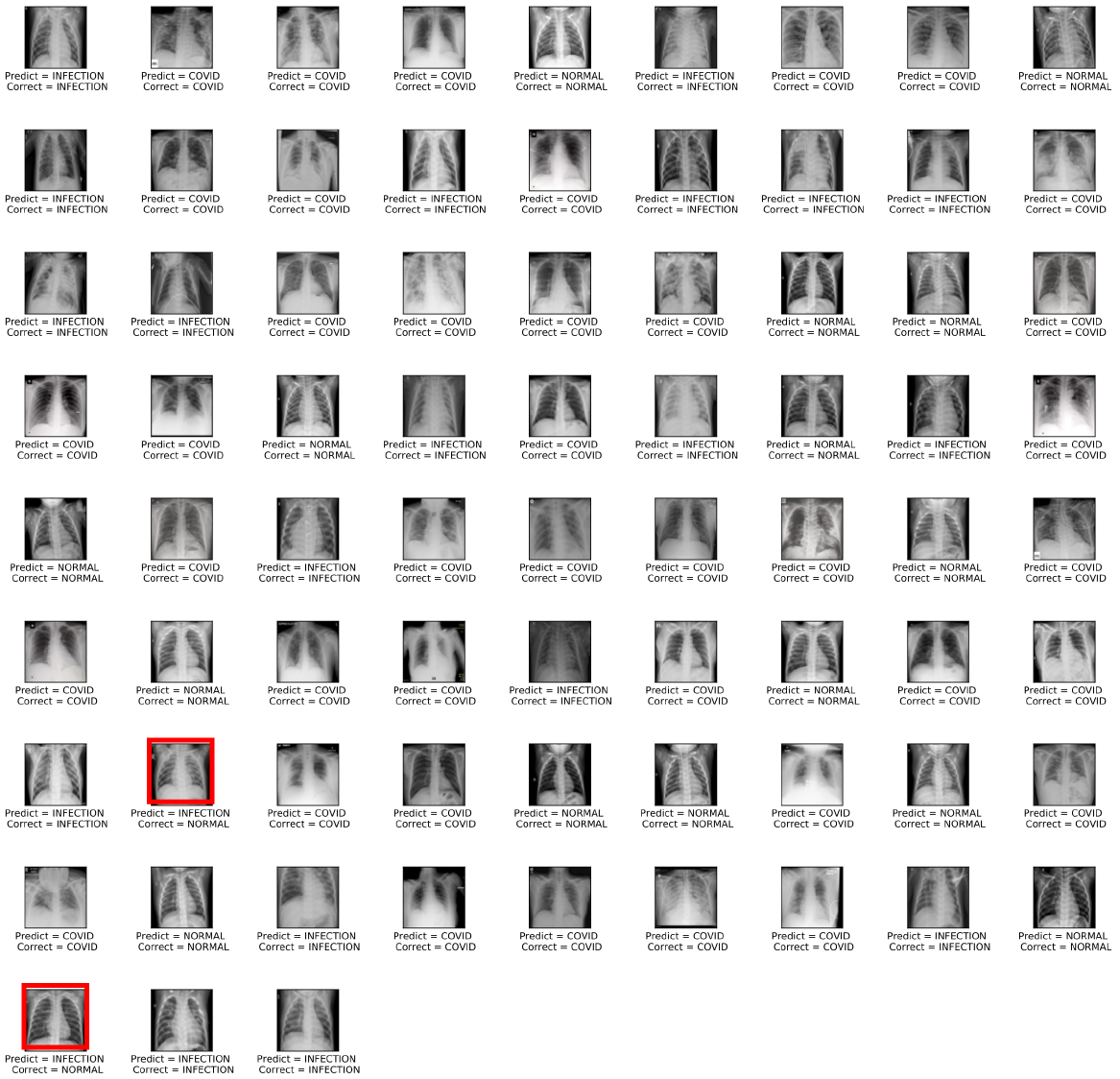}
  \caption{Results of the VGG-16 network.}
  \label{fig:covid_pulmao}
\end{figure}
}

The plot above (figure \ref{fig:covid_pulmao}) showed that the model had high accuracy in classifying the lungs with COVID-19. The class activation mapping below (figure \ref{fig:class}) exposed the image locations that the model used to perform the classifications.

It is possible to observe in each image group (NORMAL, COVID, and INFECTION - other lung infections from virus and bacteria) the most important regions for the model to detect each class. The “hotter” (color yellow), the greater the degree of importance of the image.

{\setlength\intextsep{0pt}
\begin{figure} [H]
  \centering
  \includegraphics[width=350pt, keepaspectratio]{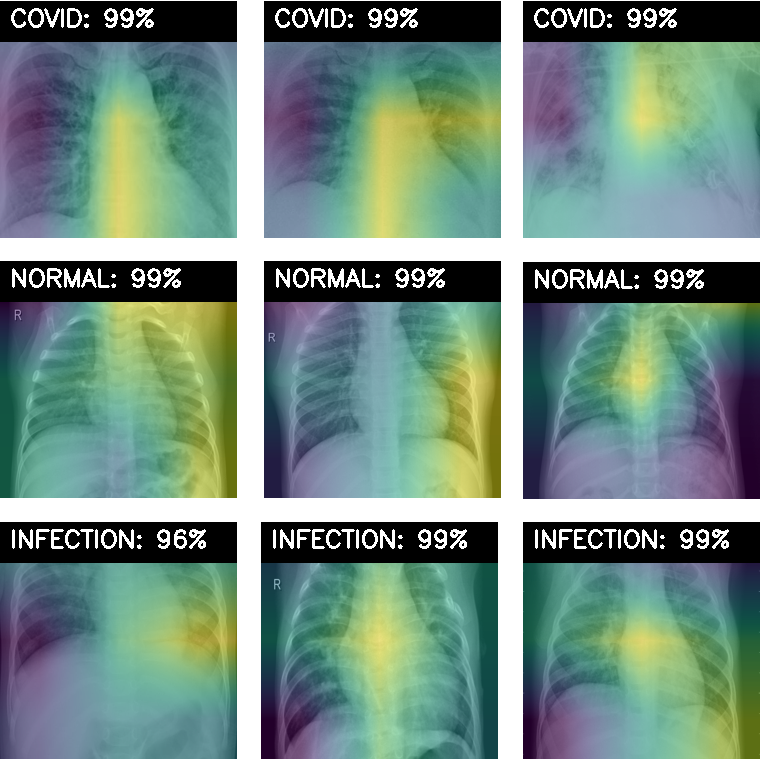}
  \caption{Class activation mapping.}
  \label{fig:class}
\end{figure}
}

\section{Discussion and Conclusion}

From the results, it is possible to notice that the model had high accuracy in classifying the healthy lungs, COVID-19, and other infections, especially using the VGG-16 architecture. Nevertheless, the class activation map did not show a clear visual pattern that allows us to infer which part of the lung the model is basing on to claim that it has COVID-19. This fact indicates the need to improve the models presented here.

Therefore, future researches may evaluate models with different architectures, parameters, and datasets that use augmentation techniques. Besides, considering the unknown character of COVID-19, the activation map exposed in this research can contribute to the literature on the possible indicators of COVID-19 in the lung images of infected individuals.

Finally, the authors highlight the importance of providing images of individuals infected with COVID-19 and also the stage of the disease. This data can contribute to the improvement of models and progress in coping with the virus. All models and codes used in this work were made available on the author's blog\footnote{\url{https://deepdados-en.github.io/}}.

\bibliographystyle{unsrt}  

%
\bibliography{references}

\end{document}